\documentclass[11pt]{article}
\usepackage[english]{babel}
\usepackage[utf8]{inputenc}
\usepackage[nottoc]{tocbibind}
\usepackage{graphicx}
\usepackage{amsmath}
\usepackage{amssymb}
\usepackage{amsxtra}

\title{Numerical simulation of Bohr-like and Thomson-like dark atoms with nuclei}
\author{T.E. Bikbaev$^{1}$, M.Yu. Khlopov$^{1,2,3}$, A.G. Mayorov$^{1}$\\
$^{1}$ National Research Nuclear University MEPhI \\115409 Moscow, Russia\\
$^{2}$ Institute of Physics, Southern Federal University\\ Stachki 194 Rostov on Don 344090, Russia\\
$^{3}$  Université de Paris, CNRS, Astroparticule et Cosmologie,\\ F-75013 Paris, France,\\ e-mail khlopov@apc.univ-paris.fr}
\date{October 2021}

\begin{document}
\maketitle

\begin{abstract}
 The puzzles of direct dark matter searches can be solved in the scenario of dark atoms, which bind hypothetical, stable, lepton-like particles with charge $-2n$, where $n$ is any natural number, with $n$ nuclei of primordial helium.
 Avoid experimental discovery because they form with primary helium neutral atom-like states $OHe$ ($X$~-- helium), called "dark" atoms. The proposed solution to this problem involves rigorous proof of the existence of a low-energy bound state in the dark atom interaction with nuclei. It implies self-consistent account for nuclear attraction and Coulomb repulsion in such an interaction. We approach the solution of this problem by numerical modeling to reveal the essence of the processes of dark atom interaction with nuclei. We start with the classical three-body problem, to which the effects of quantum physics are added. The numerical model of the dark atom interaction was developed for $O^{--}$ having a charge of -2, bound with He in Bohr-like $O$He dark atom and for $-2n$ charged $X$ bound with $n$ $\alpha$-particle nucleus in the Thomson-like atom $X$He. The development of our approach should lead to the solution of the puzzles of direct dark matter searches in the framework of dark atom hypothesis.
\end{abstract}



\section{"Dark" atoms $X$He}
If dark matter consists of particles, then they are predicted beyond the Standard Model. In particular, it is assumed that stable, electrically charged particles can exist \cite{KHLOPOV_2013, Bertone_2005, scott2011searches}.
Stable negatively charged particles can only have a charge of $-2$ ~-- we will denote them by $O^{--}$ (in the general case $-2n$, where $n$ is any natural number, we will denote them by $X$) \cite{bulekov2017search}.
In this paper, we investigate a composite dark matter scenario \cite{Kh_2008,Kh_2013,Kh_2011}. 

Hypothetical stable $O^{--}$($X$) particles avoid experimental discovery because they form neutral atom-like states $OHe$ ($X$~--helium) with primordial helium called "dark" atoms \cite{khlopov2019conspiracy}. Since all these models also predict the corresponding $+2n$ charged antiparticles, the cosmological scenario should provide a mechanism for their suppression, which, naturally, can take place in the charge-asymmetric case, corresponding to an excess of $-2n$ charged particles \cite{KHLOPOV_2013}. The electric charge of the excess of these particles is compensated by the corresponding excess of positively charged baryons. So the electroneutrality of the Universe is preserved. Hence, positively charged antiparticles can effectively annihilate in the early universe. There are various models predicting such stable $-2n$ charged particles \cite{belotsky2006composite,Khlopov_2006,Khlopov_2008}.

A "dark" atom is a system consisting of $-2n$ charged particles (in the case $n = 1$, this is $O^{--}$), bound by the Coulomb force with $n$ $^4$He nuclei. The structure of bound state depends on the value of $a \approx Z_{\alpha} Z_o \alpha A m_p R_{nHe}$ parameter, where $\alpha$ is fine structure constant, $Z_o$ and $Z_{\alpha}$--~ are the charge numbers of particle $X$ and $n$ nuclei of $He$, respectively, $m_{p}$ --~  is the proton mass, $A$ is the mass number of $n$~--nucleus $He$, and $R_{nHe}$ is the radius of the corresponding nucleus.

For $0 < a < 2$, the bound state looks like a Bohr atom with a negatively charged particle in the core and a nucleus moving in a Bohr orbit. For $2 <a< \infty$, the bound states look like Thomson's atoms, in which the body of the nucleus vibrates around a heavy negatively charged particle.

In model of $X$--~helium, $X$ behaves like a lepton or as a specific cluster of heavy quarks of new families with suppressed hadron interaction \cite{khlopov2005composite}. And the experimental lower limit on the mass of multiple charged stable particles is about $1 \text{TeV}$ \cite{beylin2020new}.


The main problem with $X$He atoms is their strong interaction with matter. This is because $X$--~helium has an unshielded nuclear attraction to the nuclei of matter. This, in turn, can lead to the destruction of a bound system of dark matter atoms and the formation of anomalous isotopes. 
To avoid the problem of overproduction of anomalous isotopes, it is necessary that the effective potential between $XHe$ and the nucleus of matter has a barrier preventing the fusion of $He$ and/or $X$ with the nucleus. In this paper, we construct a numerical model of such an interaction, to carry out calculations and calculate the interaction potential.

\section{Improvement of description of the interaction of a "dark" atom in Bohr's model with the nucleus of matter by adding the Stark effect.}

For $0 < a < 2$, the bound state of a dark atom looks like a Bohr atom. That is, dark atoms are $OHe$ atoms with $Z_{\alpha} = 2$ and $Z_o=-2$. The process of constructing a numerical model of the interaction of $OHe$ with the nucleus and the results of this interaction are given in article \cite{bikbaev2020numerical}. In this section, we have improved this numerical model by adding the Stark effect to it.

We fix the $He$ rotation orbit in the $OHe$ atom, which excludes the possibility of its polarization and we observe the Coulomb repulsion. On the other hand, the Stark effect should take place in the external electric field of a target nucleus, which leads to polarization of $OHe$. In our semiclassical numerical model, this can be taking into account by including the interaction dipole moment $\delta$ caused by the Stark effect. Thus, by manually including $\delta$, we calculated the Stark force, which is obtained from the potential and is specified using the same dipole moment.
$\delta$ appears due to the action of the nuclear force and the Coulomb force on the $He$ nucleus and also the Coulomb force on $O^{--}$, from here you can get the expression for $\delta$:
\begin{equation}
    \delta(\vec{r})=\cfrac{Z_{\alpha}E(\vec{r})}{Z_o\rho}+\cfrac{|\Vec{F_{i}}^{N}_{\alpha}|}{e\rho Z_o},
    \label{eq}
\end{equation}
where $E$ is the strength of the external electric field, $\rho=\cfrac{Z_{\alpha}e}{R_{b}^3}$ is the charge density of the $He$ nucleus, where $R_{b}$ is the Bohr radius of $He$ rotation in “dark” $O$He atoms and $|\Vec{F_{i}}^{N}_{\alpha}|$ is nuclear interaction of the Saxon-Woods type, between the $He$ nucleus and the target nucleus \cite{bikbaev2020numerical}.

The Stark potential is calculated as follows: $U_{St}=e Z_{\alpha} E \delta $. And the Stark force, respectively: $\Vec{F}_{St}=-\operatorname{grad}U_{St}$.

Bearing in mind the problem of interpreting the results of the $DAMA/NaI$ experiment on the direct search for dark matter atoms, we concentrate our calculations on the case when the target nucleus is $Na$ \cite{Kh_2011}. Therefore, in all subsequent pictures, the target nucleus should be understood as the nucleus $Na$.

Based on the data obtained, the program builds the trajectories of the $\alpha$~--particle and the $O^{--}$ particle (see Figures 1 and 2). In Figures 1 and 2, showing the result of the program, the black circle shows the location of the target nucleus, the blue dots and the red dotted line show the trajectories of $\alpha$~-- particles and $O^{--}$ particles in the XY plane, respectively.
\begin{figure}[h!]
\centering
\includegraphics[scale=0.45]{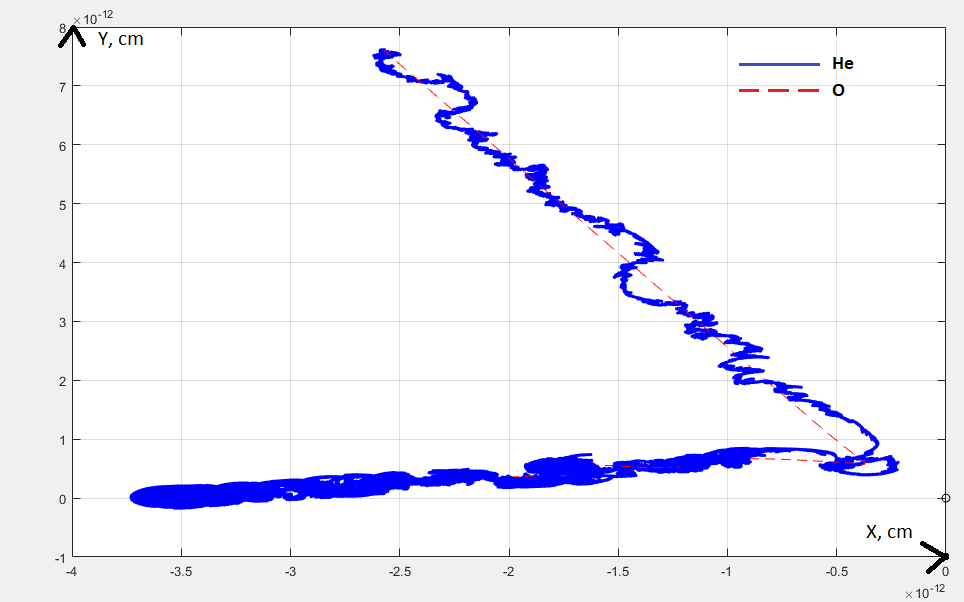}
\caption{Trajectories of $\alpha$~--particle and $O^{--}$ particle}
\label{fig:lol2}
\end{figure}

\begin{figure}[h!]
\centering
\includegraphics[scale=0.45]{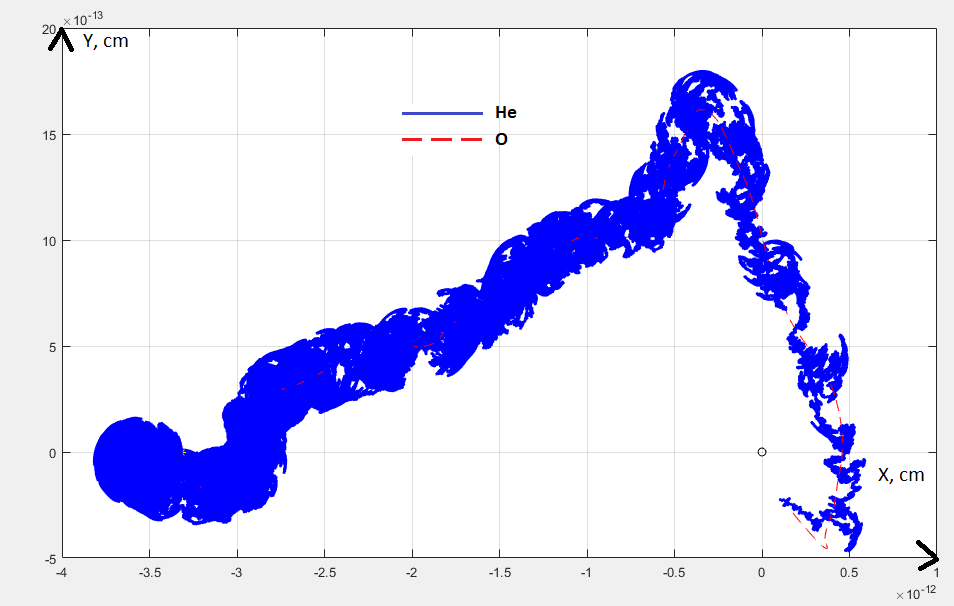}
\caption{Trajectories of $\alpha$~--particle and $O^{--}$ particle}
\label{fig:lol2}
\end{figure}

For the corresponding trajectories of the $\alpha$~--particle, it is possible to construct the total interaction potential between $He$ and the target nucleus depending on the distance between $He$ and the target nucleus (see Figures 3 and 4).
\begin{figure}[h!]
\centering
\includegraphics[scale=0.37]{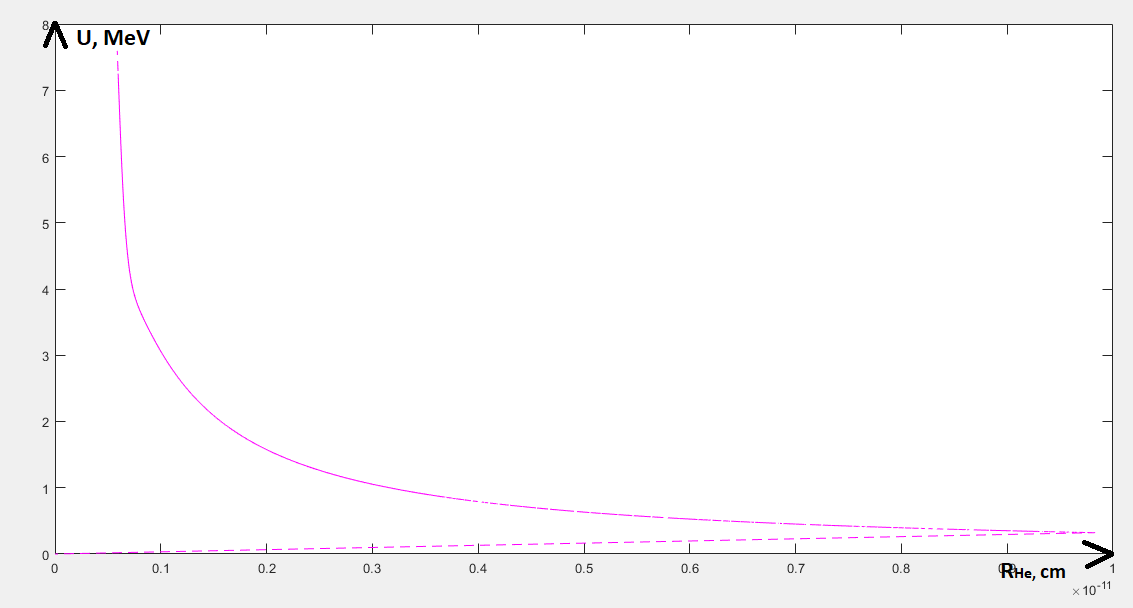}
\caption{Total potential of interaction between $He$ and target nucleus}
\label{fig:lol2}
\end{figure}

\begin{figure}[h!]
\centering
\includegraphics[scale=0.37]{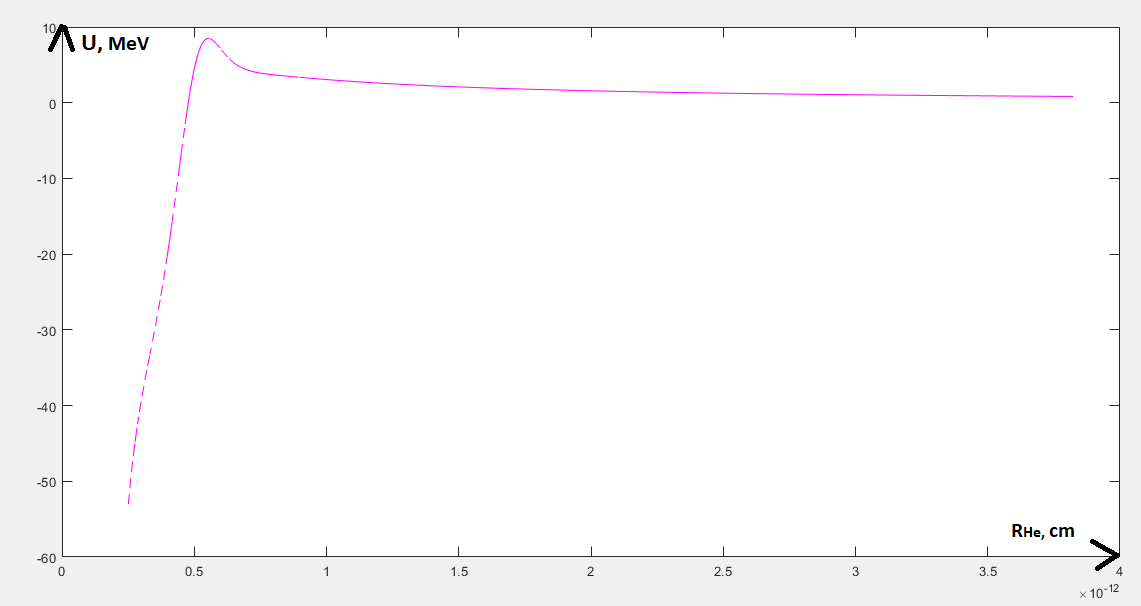}
\caption{Total potential of interaction between $He$ and target nucleus}
\label{fig:lol2}
\end{figure}

In Figure 1, a dark atom $OHe$ was scattered by a target nucleus of matter. This also follows from Figure 3, where you can see the Coulomb barrier preventing the particles of a dark atom from entering the nucleus.

In Figure 2, approximately at coordinates $(0,1; -2,5)$, the trajectories of the particles are interrupted, because the dark atom is destroyed and falls into the target nucleus. This is confirmed in Figure 4, where one can see the predominance of the nuclear potential over the Coulomb potential at distances close to the target nucleus.

The results of the interaction can be quite varied, which requires a detailed study by collecting statistics of trajectories with varying the initial values of the system and the parameters of the target nucleus.

The approach of the Bohr atom model has some drawbacks, for example, in our numerical model, the Coulomb force between helium and $O^{--}$ is not explicitly specified, but the $He$ rotation orbit in the $OHe$ atom is manually fixed, which excludes the possibility of its polarization. And when considering Thomson's model of the atom, this problem can be solved, since with this approach helium is not a point charge stochastically moving in a fixed Bohr orbit, but is a charged ball inside which the particle $O^{--}$ can oscillate. Moreover, the case of $-2$ charged particles is only a special case, since the particles we are considering can have a charge $-2n$ and form with $n$ nuclei $^4$He "dark" atoms $X$~--helium, which by themselves, starting from $n=2$, are Thomson atoms. With all this, the Stark effect, when considering the Thomson atom, should arise by itself automatically.

It should be clarified that in the following sections we simulated a dark $XHe$ atom and its interaction with a target nucleus, and in all the figures the case of a Thomson atom at $n=1$ was considered. That is, we considered the $OHe$ atom in the Thomson approximation. But since $OHe$ is a special case of $XHe$, and our model successfully describes the general case, we left the designation $XHe$ in all the following sections.

\section{Numerical simulation of the interaction of Thomson's "dark" atom with the nucleus.}
\subsection{Modeling $X$~-- helium}
The "dark" atom of $X$ ~-- helium is the bound state of an n~--$\alpha$-particle (n~-- helium) nucleus and the particle $X$ with charge $-2n$. We place the spherical coordinate system at the center of the $n$~--helium nucleus, which is a charged ball. Inside which, in the center, there is a point particle $X$. When external forces begin to act, the distance between the center of $n$~--helium and $X$ becomes nonzero and the particle $X$ begins to oscillate inside the $nHe$ nucleus (in reality, $nHe$ is much lighter than $X$, therefore it is a nuclear a drop that fluctuates around $X$).

The force of the Coulomb interaction between $n$ ~--helium and $X$ is given by the following formula:
\begin{equation}
\Vec{F}_{XHe}(R_{XHe})=
\begin{cases}
-\cfrac{4e^2n^2}{R_{XHe}^3}\Vec{R}_{XHe} & \text{for}\hspace{5pt} R_{XHe}>R_{He},\\
-\cfrac{4e^2n^2}{R_{He}^3}\Vec{R}_{XHe} & \text{for}\hspace{5pt} R_{XHe}<R_{He},
\end{cases}
\end{equation}
where $|\vec{R}_{XHe}|$ is the distance between $X$ and the center of the $nHe$ nucleus, and $ R_{He}$ is the radius of the $n$~--helium nucleus.

The scheme of numerical simulation of the dynamical system $XHe$:

1) Initial coordinates $X$ $\Vec{R_0}_{X} = 0$ and its initial speed, which we set equal to the thermal speed in the medium, $V_{0_{X}}=\left(\cfrac{3 k T}{M_{n u c}}\right)^{1/2}$, where $M_{nuc}$ is the mass of the target nucleus, $T$ is the temperature (we take 25 degrees Celsius), and $k$ is the Boltzmann constant.

2) Consider state of the system at next moment of time, taken on the time interval dt. The i-th value of increment of components of radius vector $X$ is determined, $dr_i$:
\begin{equation}
    dr_i=V_{i_{X}}dt.
    \label{eq}
\end{equation}

3) The i + 1 value of the components of the radius vector $X$ is calculated, $r_{i+1}$:
\begin{equation}
    r_{i+1}=r_i + dr_i.
    \label{eq}
\end{equation}

4) In each iteration, the program calculates the force acting on $X$, $\Vec{F_i}_{XHe}$. Using which the increment of the momentum  $d\Vec{P_{i}}$ of the particle $X$ is determined:
\begin{equation}
    d\Vec{P_{i}}=\Vec{F_i}_{XHe}dt.
    \label{eq}
\end{equation}

5) Using the increment of the momentum $d\Vec{P_{i}}$, the increment of the particle velocity $X$, $d\Vec{V_{i_X}}$, is calculated, for the subsequent finding of the new velocity used in the next iteration:
\begin{equation}
    d\Vec{V}_{i_X}=\cfrac{d\Vec{P_{i}}}{m_{X}}.
    \label{eq}
\end{equation}

Using the obtained data, it is possible to plot the dependence of modulus of the radius vector of particle $X$ on time (see Figure 5). In Figure 5 one can observe the oscillation of the particle $X$ inside the nucleus $nHe$ with a period approximately equal to $2\cdot10^{-20}$ second. They appear because the Coulomb force between the nucleus $nHe$ and $X$ tends to return $X$ to the center of the nucleus and to neutralize the external disturbance given to the particle $X$. $R_{XHe} < 1 \hspace{1.5mm}\text{fm}$, which indicates the stability of the $X$~-- helium system.
\begin{figure}[h!]
\centering
\includegraphics[scale=0.4]{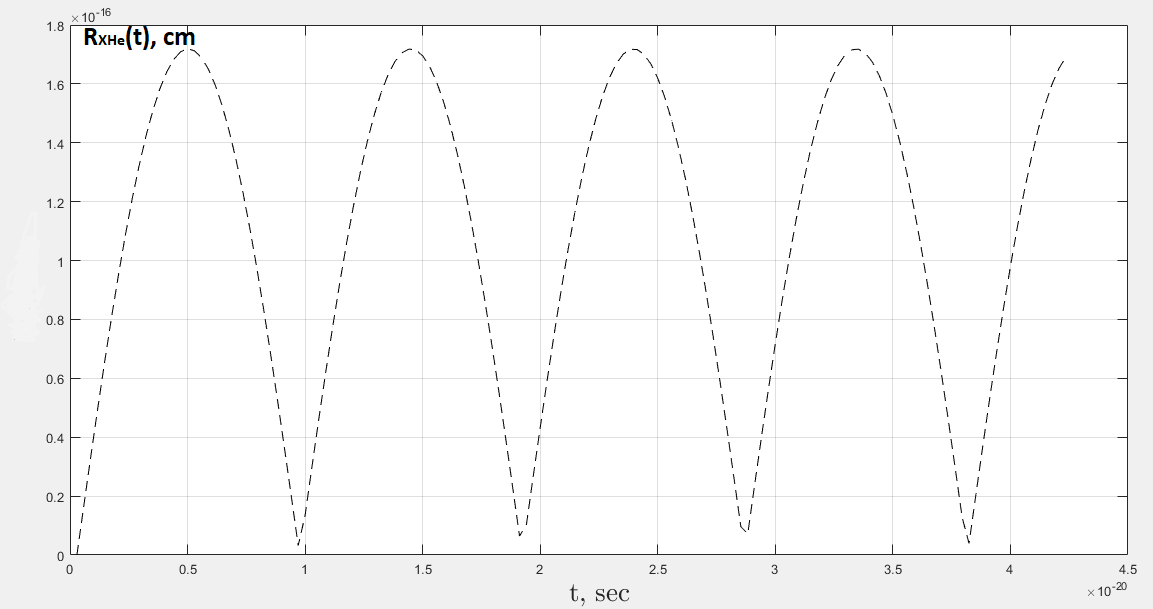}
\caption{Dependence of modulus of radius vector of particle $X$ on time t}
\label{fig:lol1}
\end{figure}

\subsection{Interaction in the $XHe$~--nucleus system}
The coordinate system $XHe$ ~-- is the core, in which the interaction of $XHe$ with the target nucleus will be simulated, similar to the coordinate system $OHe$ ~-- the core described in the 3.2 paragraph of article \cite{bikbaev2020numerical}. The difference is that the distance between $X$ and $nHe$ is no longer strictly fixed and is not equal to the Bohr radius. Thus, the radius vector $nHe$, $r_{He}$, and $X$, $r$, are determined independently, and the distance between $X$ and $nHe$, $r_{XHe} $, is determined as follows:
\begin{equation}
    \vec{r}_{XHe}=\vec{r}_{\alpha}-\vec{r}
    \label{eq}
\end{equation}

In our Thomson approximation, helium is a charged droplet, but when considering the coulomb force and the nuclear force between $He$ and the target nucleus, non-point of helium has not yet been taken into account and this will have to be done in the future.

Therefore, the Coulomb and nuclear forces acting between the particles of a dark atom and the target nucleus in the $XHe$~--nucleus system are similar to the forces described in paragraphs 3.3 and 3.4 of article \cite{bikbaev2020numerical}.
To these forces are added two additional forces, equal in magnitude, but opposite in sign. This is the Coulomb force between $X$ and $nHe$ (see formula 2). The force acting on $ nHe $ is denoted by $\Vec{F_{i}}^{XHe}_{\alpha}$. And the force acting on $X$ is denoted by $\Vec{F_{i}}^{XHe}_{X}=-\Vec{F_{i}}^{XHe}_{\alpha}$.

The total force acting on the particle $X$, $\Vec{F_{i}}^{X}_{Sum}$, is calculated as follows:
\begin{equation}
    \Vec{F_{i}}^{X}_{Sum}=\Vec{F_{i}}^{e}_{ZO}+\Vec{F_{i}}^{XHe}_{X}.
    \label{eq}
\end{equation}

The total force acting on $nHe$, $\Vec{F_{i}}_{\alpha}$, is:
\begin{equation}
    \Vec{F_{i}}_{\alpha}=\Vec{F_{i}}^{e}_{\alpha}+\Vec{F_{i}}^{N}_{\alpha} + \Vec{F_{i}}^{XHe}_{\alpha}.
    \label{eq}
\end{equation}

Let us construct a numerical scheme for calculating these forces depending on the distance between objects.

1) We use the following initial conditions: initial coordinates $X$ and $nHe$, $\Vec{r_0}=\Vec{r_0}_{\alpha}$, and their initial velocities, which we set equal to the thermal speed in the medium, $V_{{X}_0}=V_{{\alpha}_0}=\left(\cfrac{3 k T}{M_{n u c}}\right)^{1/2}$. 

2) Consider the state of the system at the next moment in time, taken on the time interval dt. The i-th value of the impulse increment $nHe$, $d\Vec{P_{i}}_{\alpha}$, and $X$, $d\Vec{P_{i}}$, is determined:
\begin{equation}
    d\Vec{P_{i}}_{\alpha}=\Vec{F_{i}}_{\alpha}dt,
    \label{eq}
\end{equation}
\begin{equation}
    d\Vec{P_{i}}=\Vec{F_{i}}^{X}_{Sum}dt.
    \label{eq}
\end{equation}

3) Using the increment $d\Vec{P_{i}}_{\alpha}$ and $d\Vec{P_{i}}$, i + 1 values of the velocities of the nucleus $nHe$ and $X$, $\Vec{V}_{{\alpha}_{i+1}}$ and $\Vec{V}_{X_{i+1}}$:
\begin{equation}
    \Vec{V}_{{\alpha}_{i+1}}=\Vec{V}_{{\alpha}_{i}} + \cfrac{d\Vec{P_{i}}_{\alpha}}{m_{He}},
    \label{eq}
\end{equation}

\begin{equation}
    \Vec{V}_{X_{i+1}}=\Vec{V}_{X_{i}} + \cfrac{d\Vec{P_{i}}}{m_{X}}.
    \label{eq}
\end{equation}

4) Calculate the i + 1 value of the radius vector $X$ and $nHe$:
\begin{equation}
    \Vec{r}_{i+1}=\Vec{r}_i + \Vec{V}_{{\alpha}_{i+1}} dt,
    \label{eq}
\end{equation}

\begin{equation}
    \Vec{r}_{\alpha_{i+1}}=\Vec{r}_{\alpha_{i}} + \Vec{V}_{X_{i+1}}dt,
    \label{eq}
\end{equation}

5) In each cycle, program calculates total force acting on $X$ particle, $\Vec{F_{i}}^{X}_{Sum}$, and the total force acting on $nHe$, $\Vec{F_{i}}_{\alpha}$.

The dependence of the radius of the particle vector $X$ on the radius of the vector of the $n$ ~--helium nucleus see Figure 6 and the total potential of $nHe$ interaction with the target nucleus depending on $r_{\alpha}$ see Figure 7.
\begin{figure}[h!]
\centering
\includegraphics[scale=0.38]{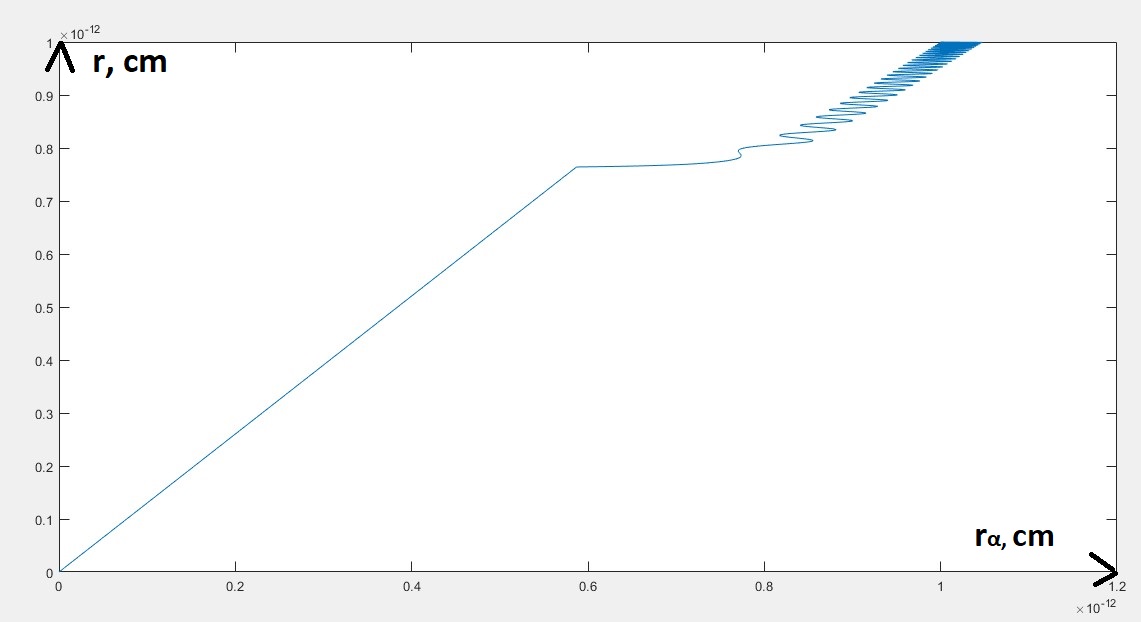}
\caption{Dependence of $r$ on $ r_{\alpha}$}
\label{fig:lol1}
\end{figure}

\begin{figure}[h!]
\centering
\includegraphics[scale=0.38]{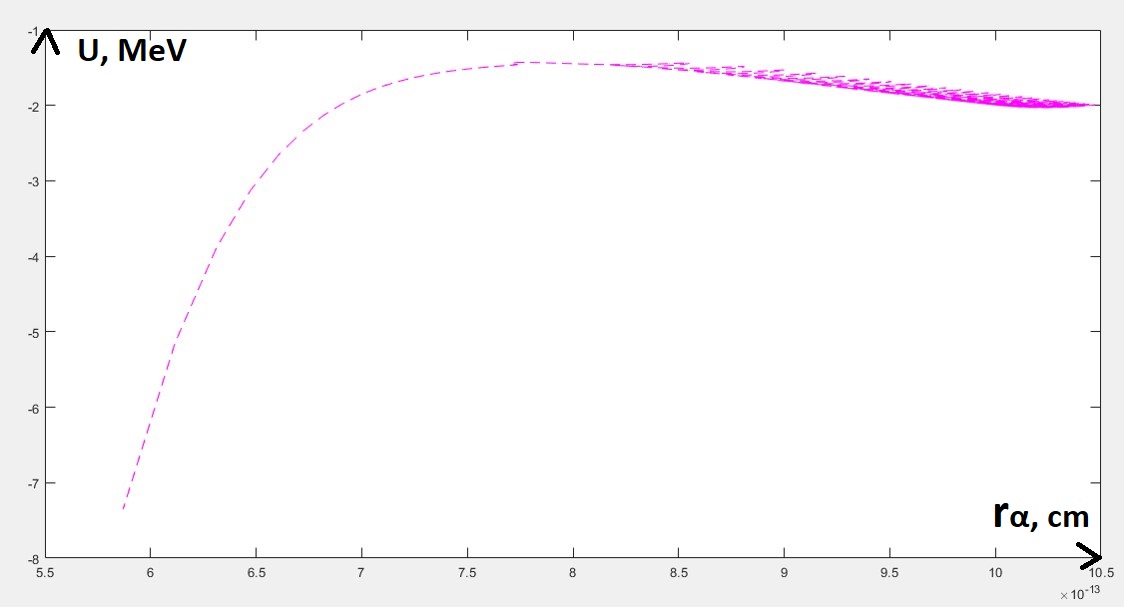}
\caption{Dependence of the total potential of $nHe$ interaction with the target nucleus on $r_{\alpha}$}
\label{fig:lol1}
\end{figure}

It can be seen from the figures that the $XHe$ system moves towards the target nucleus as a bound system. The radius vector of the $X$ particle is always less than the radius of the $He$ vector at the same time, that is, the $X$ particle is slightly closer to the target nucleus than helium (see Figure 6). Therefore, we see the polarization of the "dark" atom. But at a sufficiently close distance from the target nucleus, the nuclear force becomes strong enough to overcome the Coulomb repulsion of $nHe$ by the target nucleus and $n$~--helium, pushing forward, penetrates the nucleus, which is clearly seen in Figure 7.

After that, we supplemented the Coulomb force acting between $nHe$ and the nucleus, and the Coulomb force acting between $X$ and the nucleus, similarly to formula 2, i.e. added a condition so that its form would change upon penetration of $nHe$ and $X$ particles into the target nucleus.

The main task of our modeling is to reconstruct the total potential of interaction between $nHe$ and the target nucleus.

From the analysis of trajectories, two characteristic cases can be distinguished. With a zero impact parameter, the $ XHe $ atom flies through the target nucleus, then comes back and flies in the opposite direction (see Figures 8).

\begin{figure}[h!]
\centering
\includegraphics[scale=0.4]{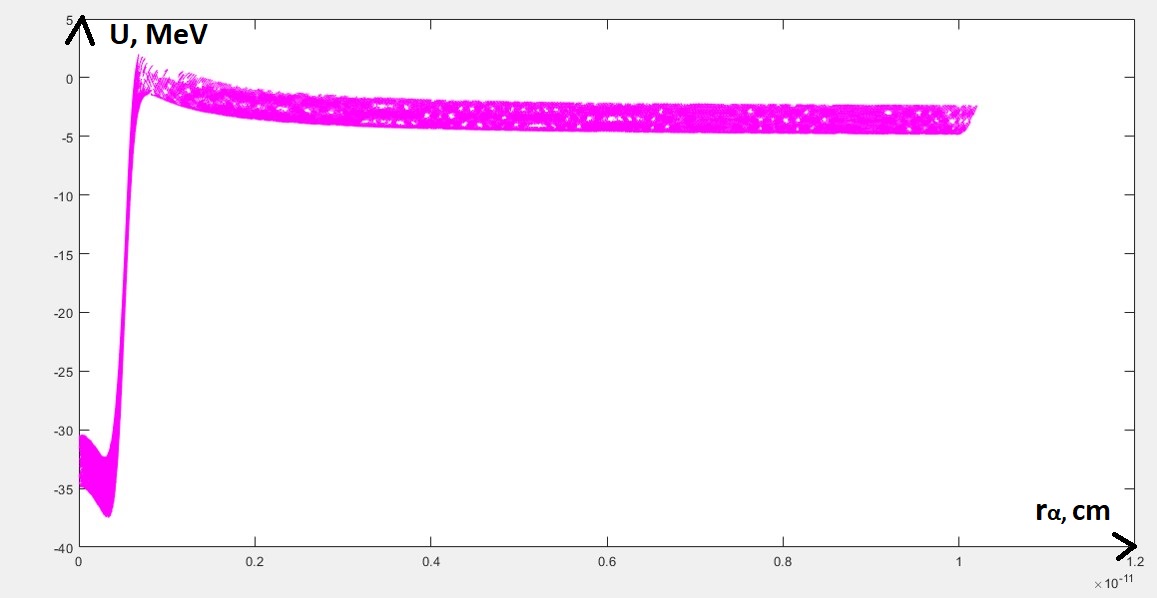}
\caption{Dependence of the total potential of $nHe$ interaction with the target nucleus on $r_{\alpha}$}
\label{fig:lol1}
\end{figure}

It is assumed that the interaction of slow $X$~--helium atoms with nuclei can lead to their low-energy binding. Thus, the low-energy bound state of the $XHe$~--nucleus must be an oscillating three-body system. And we see that with a nonzero impact parameter, the $XHe$ atom hits the target nucleus, and a certain oscillatory system of three bodies is formed (this can be seen in Figures 9 and 10).

\begin{figure}[h!]
\centering
\includegraphics[scale=0.4]{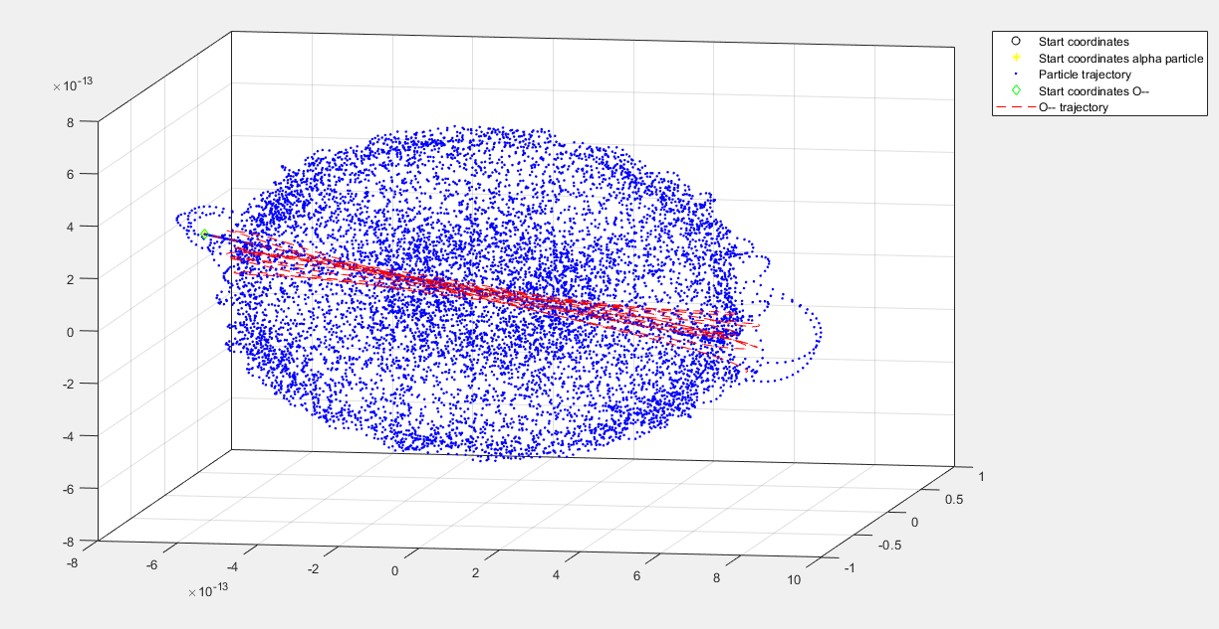}
\caption{Trajectories of $nHe$ and particle $X$}
\label{fig:lol1}
\end{figure}
In Figure 9, the black circle shows the target nucleus, the yellow asterisk and the green rhombus are the initial locations of $nHe$ and the $X$ particle, respectively, the blue dots and the red dotted line show the trajectories of $nHe$ and the $X$ particle, respectively.
\begin{figure}[h!]
\centering
\includegraphics[scale=0.4]{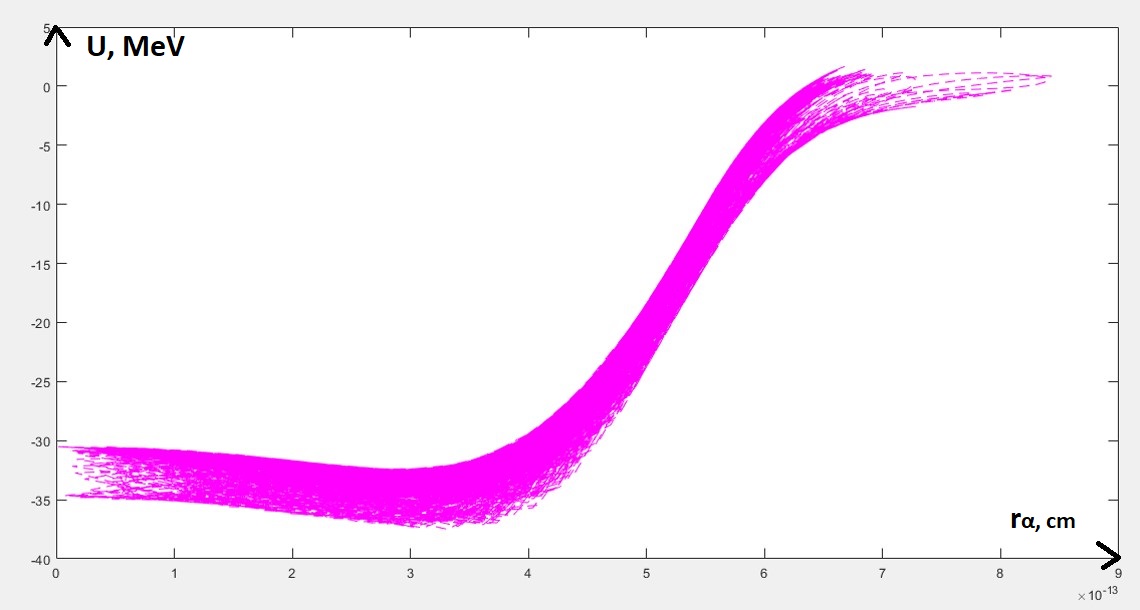}
\caption{Dependence of the total potential of $nHe$ interaction with the target nucleus on $r_{\alpha}$}
\label{fig:lol1}
\end{figure}

When analyzing the trajectories, it turned out that the variation in the mass of $X$ does not affect the result in any way. There is a dependence on the aiming parameter and the initial speed of the system. But for any of their values, the cloud of particle coordinates is inside the target nucleus. This is probably reasonable, since the zero balance of forces for helium can only be achieved in the region where the nuclear and Coulomb forces from the nucleus are balanced. But the nuclear force is small outside of it. Thus, the Coulomb polarization of $XHe$ occurs up to the nuclear boundary, and nuclear polarization becomes possible only inside it.

\section{Conclusions}

The paper investigates the hypothesis of composite dark matter, in which hypothetical stable particles with a charge of $-2n$ form neutral atom-like states $XHe$ with primary helium nuclei. $X$~--helium will interact with the nuclei of ordinary matter.
The nuclear interaction of dark atoms with matter is a key problem in the composite dark matter scenario. Solving this problem and correctly describing this interaction will reveal the role of dark atoms in primary nucleosynthesis, stellar processes, and will also explain the conflicting results of experiments on the direct search for dark matter due to the peculiarities of the interaction of “dark” atoms with the substance of underground detectors \cite{BERNABEI_2013}.

The $XHe$ hypothesis cannot work unless a repulsive interaction occurs at some distance between $XHe$ and the nucleus, and the solution of this problem is vital for the further existence of the $XHe$ atomic model of dark matter \cite{Khlopov:2015nrq}. Therefore, we were faced with the task of constructing a numerical model of the interaction of $XHe$ with a target nucleus. Such a numerical model is constructed in this work in the form of a Thomson model of the atom, as an attempt to avoid the disadvantages found in Bohr's model. Our model describes a system of three charged particles interacting with each other by means of Coulomb and nuclear forces.


When simulating in the Thomson atom approximation, the following effects were observed: with a zero impact parameter, the $XHe$ atom flies through the target nucleus, then returns and flies in the opposite direction; with a nonzero impact parameter, the $XHe$ atom hits the target nucleus, and a kind of vibrational system of three bodies, this is what is expected to be seen in the formation of a low-energy bound state in the interaction of slow $X$~--helium atoms with the nuclei of matter. However, the disadvantage of this is that particle oscillations occur inside the target nucleus. Thus, in the current version of the numerical model, $nHe$ can easily penetrate into the target nucleus and the elastic collisions of nuclei do not arise. In the approach of the Bohr atom, the opposite was true \cite{bikbaev2020numerical}. Therefore, in the future it should be taken into account that nuclear matter is incompressible and opaque. In other words, we need to figure out how to take into account the real properties of nuclear matter.

\section*{Acknowledgements}
The work by TB and AM has been supported by the grant of the Russian Science Foundation (Project No-18-12-00213-P). 




\begin{thebibliography}{99}
\bibitem{KHLOPOV_2013} M. Khlopov: Fundamental particle structure in the cosmological dark matter, International Journal of Modern Physics A. \textbf{28}, 1330042 (2013)

\bibitem{Bertone_2005} G. Bertone, D. Hooper, J. Silk: Particle dark matter: evidence, candidates and constraints, Physics Reports \textbf{405}, 279 ~-- 390 (2005) 

\bibitem{scott2011searches} P. Scott: Searches for Particle Dark Matter: An Introduction, (2011), e-Print: arXiv:1110.2757.

\bibitem{bulekov2017search} O. V. Bulekov, M.Yu.Khlopov, A. S. Romaniouk, Yu. S. Smirnov: Search for Double Charged Particles as Direct Test for Dark Atom Constituents, Bled Workshops in Physics \textbf{18}, 11-24 (2017)

\bibitem{Kh_2008} M. Yu. Khlopov, C. Kouvaris: Composite dark matter from a model with composite Higgsboson, Phys. Rev. \textbf{78}, 065040 (2008)

\bibitem{Kh_2013} D. Fargion, M. Yu. Khlopov: Tera-leptons’ shadows over Sinister Universe, Gravitation Cosmol. \textbf{19}, 219 (2013)

\bibitem{Kh_2011} M. Yu. Khlopov, A. G. Mayorov, and E. Yu. Soldatov: Towards nuclear physics of OHe darkmatter, Bled Workshops Phys. \textbf{12}, 94 (2011)

\bibitem{khlopov2019conspiracy} M. Y. Khlopov: Conspiracy of BSM physics and cosmology, Bled Workshops in Physics, V.20 PP.21-35 (2019), e-Print: arXiv: 1911.03294.

\bibitem{belotsky2006composite} K. M. Belotsky, M. Y. Khlopov, K. I. Shibaev: Composite Dark Matter and its Charged Constituents, Grav.Cosmol., V.12 PP.93-99, (2006), arXiv:astro-ph/0604518

\bibitem{Khlopov_2006} M. Y. Khlopov, C. A. Stephan, D. Fargion: Dark matter with invisible light from heavy double charged leptons of almost-commutative geometry?, Classical and Quantum Gravity \textbf{23}, 7305~--7354 (2006)

\bibitem{Khlopov_2008} M. Y. Khlopov, C. Kouvaris: Strong interactive massive particles from a strong coupled theory, Physical Review D \textbf{77}, PP. 065002 (2008)

\bibitem{khlopov2005composite} M. Y. Khlopov: Composite dark matter from 4th generation, JETP Letters \textbf{83}, 1~--4 (2006)

\bibitem{beylin2020new} V. Beylin, M. Khlopov, V. Kuksa, N. Volchanskiy: New physics of strong interaction and Dark Universe, Universe \textbf{6}, 196 (2020)

\bibitem{bikbaev2020numerical} T. E. Bikbaev, M. Yu. Khlopov, A. G. Mayorov: Numerical simulation of dark atom interaction with nuclei, Bled Workshops in Physics \textbf{21}, 105~--117 (2020)

\bibitem{BERNABEI_2013} R. Bernabei: Dark matter investigation by DAMA in Gran Sasso, International Journal of Modern Physics A \textbf{28}, 1330022 (2013)

\bibitem{Khlopov:2015nrq} M. Yu. Khlopov: 10 years of dark atoms of composite dark matter, Bled Workshops Physics \textbf{16}, 71~--77 (2015)







\bibitem{Cudell:2012fw} J. R. Cudell, M. Y. Khlopov, Q. Wallemacq: The nuclear physics of OHe, Bled Workshops Physics \textbf{13}, 10~--27 (2012)





\end{thebibliography}
\end{document}